\documentclass[aps,prl,reprint,superscriptaddress,nofootinbib, showpacs]{revtex4-2}
\preprint{FERMILAB-PUB-26-0335-PPD}
\usepackage[absolute,overlay]{textpos}
\usepackage{graphicx}% Include figure files
\usepackage{dcolumn}% Align table columns on decimal point
\usepackage{amsmath}
\usepackage{bm}% bold math
\usepackage{orcidlink}
\usepackage{hyperref}
\hypersetup{
    colorlinks=true,
    linkcolor=blue,
    citecolor=blue,
    urlcolor=blue
}

\begin{document}
\begin{textblock*}{5cm}(14.5cm,1.5cm)
\raggedleft\small FERMILAB-PUB-26-0335-PPD
\end{textblock*}

\title{First Measurement of Sub-GeV $\nu_{\mu}$ Charged-Current Coherent Pion Production on Argon in MicroBooNE}% 

%\date{\today}

\newcommand{\ANL}{Argonne National Laboratory (ANL), Lemont, IL 60439, USA}
\newcommand{\Bern}{Universit\"{a}t Bern, Bern CH-3012, Switzerland}
\newcommand{\BNL}{Brookhaven National Laboratory (BNL), Upton, NY 11973, USA}
\newcommand{\UCSB}{University of California, Santa Barbara, CA 93106, USA}
\newcommand{\Cambridge}{University of Cambridge, Cambridge CB3 0HE, United Kingdom}
\newcommand{\CIEMAT}{Centro de Investigaciones Energ\'{e}ticas, Medioambientales y Tecnol\'{o}gicas (CIEMAT), Madrid E-28040, Spain}
\newcommand{\Chicago}{University of Chicago, Chicago, IL 60637, USA}
\newcommand{\Cincinnati}{University of Cincinnati, Cincinnati, OH 45221, USA}
\newcommand{\CSU}{Colorado State University, Fort Collins, CO 80523, USA}
\newcommand{\Columbia}{Columbia University, New York, NY 10027, USA}
\newcommand{\Edinburgh}{University of Edinburgh, Edinburgh EH9 3FD, United Kingdom}
\newcommand{\FNAL}{Fermi National Accelerator Laboratory (FNAL), Batavia, IL 60510, USA}
\newcommand{\Granada}{Universidad de Granada, Granada E-18071, Spain}
\newcommand{\IIT}{Illinois Institute of Technology (IIT), Chicago, IL 60616, USA}
\newcommand{\ICL}{Imperial College London, London SW7 2AZ, United Kingdom}
\newcommand{\Indiana}{Indiana University, Bloomington, IN 47405, USA}
\newcommand{\Kansas}{The University of Kansas, Lawrence, KS 66045, USA}
\newcommand{\KSU}{Kansas State University (KSU), Manhattan, KS 66506, USA}
\newcommand{\Lancaster}{Lancaster University, Lancaster LA1 4YW, United Kingdom}
\newcommand{\LANL}{Los Alamos National Laboratory (LANL), Los Alamos, NM 87545, USA}
\newcommand{\Louisiana}{Louisiana State University, Baton Rouge, LA 70803, USA}
\newcommand{\Manchester}{The University of Manchester, Manchester M13 9PL, United Kingdom}
\newcommand{\MIT}{Massachusetts Institute of Technology (MIT), Cambridge, MA 02139, USA}
\newcommand{\Michigan}{University of Michigan, Ann Arbor, MI 48109, USA}
\newcommand{\MSU}{Michigan State University, East Lansing, MI 48824, USA}
\newcommand{\Minnesota}{University of Minnesota, Minneapolis, MN 55455, USA}
\newcommand{\Nankai}{Nankai University, Nankai District, Tianjin 300071, China}
\newcommand{\NMSU}{New Mexico State University (NMSU), Las Cruces, NM 88003, USA}
\newcommand{\Oxford}{University of Oxford, Oxford OX1 3RH, United Kingdom}
\newcommand{\Pitt}{University of Pittsburgh, Pittsburgh, PA 15260, USA}
\newcommand{\QMUL}{Queen Mary University of London, London E1 4NS, United Kingdom}
\newcommand{\Rutgers}{Rutgers University, Piscataway, NJ 08854, USA}
\newcommand{\SLAC}{SLAC National Accelerator Laboratory, Menlo Park, CA 94025, USA}
\newcommand{\SDSMT}{South Dakota School of Mines and Technology (SDSMT), Rapid City, SD 57701, USA}
\newcommand{\Maine}{University of Southern Maine, Portland, ME 04104, USA}
\newcommand{\TelAviv}{Tel Aviv University, Tel Aviv 69978, Israel}
\newcommand{\UTA}{University of Texas, Arlington, TX 76019, USA}
\newcommand{\Tufts}{Tufts University, Medford, MA 02155, USA}
\newcommand{\VTech}{Center for Neutrino Physics, Virginia Tech, Blacksburg, VA 24061, USA}
\newcommand{\Warwick}{University of Warwick, Coventry CV4 7AL, United Kingdom}
\newcommand{\NotreDame}{University of Notre Dame, Notre Dame, IN 46556, USA}

\affiliation{\ANL}
\affiliation{\Bern}
\affiliation{\BNL}
\affiliation{\UCSB}
\affiliation{\Cambridge}
\affiliation{\CIEMAT}
\affiliation{\Chicago}
\affiliation{\Cincinnati}
\affiliation{\CSU}
\affiliation{\Columbia}
\affiliation{\Edinburgh}
\affiliation{\FNAL}
\affiliation{\Granada}
\affiliation{\IIT}
\affiliation{\ICL}
\affiliation{\Indiana}
\affiliation{\Kansas}
\affiliation{\KSU}
\affiliation{\Lancaster}
\affiliation{\LANL}
\affiliation{\Louisiana}
\affiliation{\Manchester}
\affiliation{\MIT}
\affiliation{\Michigan}
\affiliation{\MSU}
\affiliation{\Minnesota}
\affiliation{\Nankai}
\affiliation{\NMSU}
\affiliation{\NotreDame}
\affiliation{\Oxford}
\affiliation{\Pitt}
\affiliation{\QMUL}
\affiliation{\Rutgers}
\affiliation{\SLAC}
\affiliation{\SDSMT}
\affiliation{\Maine}
\affiliation{\TelAviv}
\affiliation{\UTA}
\affiliation{\Tufts}
\affiliation{\VTech}
\affiliation{\Warwick}

% Authors in alphabetical order:
\author{P.~Abratenko\,\orcidlink{0000-0001-6945-5941}}\affiliation{\Tufts}
\author{D.~Andrade~Aldana\,\orcidlink{0009-0008-3143-3374}}\affiliation{\IIT}
\author{J.~Asaadi\,\orcidlink{0000-0001-6915-5279}}\affiliation{\UTA}
\author{A.~Ashkenazi\,\orcidlink{0000-0002-1995-3851}}\affiliation{\TelAviv}
\author{S.~Balasubramanian}\affiliation{\FNAL}
\author{B.~Baller\,\orcidlink{0000-0001-8731-9281}}\affiliation{\FNAL}
\author{A.~Barnard\,\orcidlink{0000-0001-6117-1768}}\affiliation{\Oxford}
\author{G.~Barr\,\orcidlink{0000-0002-9763-1882}}\affiliation{\Oxford}
\author{D.~Barrow\,\orcidlink{0000-0001-5844-709X}}\affiliation{\Oxford}
\author{J.~Barrow\,\orcidlink{0000-0002-3993-1123}}\affiliation{\Minnesota}
% ...
\collaboration{The MicroBooNE Collaboration}
\thanks{microboone\_info@fnal.gov}\noaffiliation

\begin{abstract}
We report a measurement of the charged-current coherent pion production cross section on argon using the MicroBooNE liquid argon time projection chamber exposed to the Booster Neutrino Beam at Fermilab. The measurement uses the MicroBooNE data set corresponding to $1.26 \times 10^{21}$ protons on target with a mean neutrino energy of $0.8$~GeV. The flux-averaged cross section is measured to be $(9.1 \pm 1.2_{\text{stat}} \pm 1.2_\text{syst}) \times 10^{-40}\,\text{cm}^2/\text{Ar}$. This result represents the first measurement of charged-current coherent pion production on argon at sub-GeV neutrino energies. Due to its clean two-body kinematics, where the neutrino interacts coherently with the entire nucleus producing a forward muon and pion with no nuclear breakup, this process provides a useful tool for constraining neutrino flux uncertainties in current and future oscillation experiments such as DUNE.
\end{abstract}

\maketitle

%\begingroup
%\collaboration{The MicroBooNE Collaboration}
%\thanks{microboone_info@fnal.gov}\noaffiliation

%\renewcommand\thefootnote{}
%\footnotetext{*Contact author: microboone\_info@fnal.gov}
%\endgroup

\textit{Introduction ---} A precise understanding of the neutrino flux is essential for the next generation of long-baseline oscillation experiments such as the Deep Underground Neutrino Experiment (DUNE), which aims to measure neutrino oscillation parameters with unprecedented precision \cite{Abi2020DUNE}. Achieving this precision requires percent-level control of flux-related systematic uncertainties. $\nu_\mu$-induced charged-current coherent pion ($\nu_\mu$CC$\pi^+_c$) production offers a promising handle on the $\nu_\mu$ component of the neutrino flux. This follows from its simple two-body final-state kinematics, characterized by a forward muon-pion system with the nucleus remaining intact, enabling a precise determination of the neutrino energy. Due to this unique attribute, this process has been proposed as a handle for constraining neutrino fluxes in situ \cite{Jung2025}. The use of this process as a flux constraint, however, depends on an accurate understanding of its cross section motivating precise experimental measurements. Although $\nu_\mu$CC$\pi^+_c$ production is theoretically connected to pion-nucleus elastic scattering via Adler’s relation \cite{Adler}, this connection is exact only in the high-energy limit where lepton-mass effects can be neglected, and additional assumptions are required when extending these models to lower neutrino energies \cite{PhysRevD.79.013002}. Consequently, current generator predictions exhibit significant differences. Measurements of $\nu_\mu$CC$\pi^+_c$ production on argon at current liquid-argon time projection chamber (LArTPC) experiments are, therefore, critical to assess these models and to establish the viability of this channel as a flux constraint for DUNE.

In $\nu_\mu$CC$\pi^+_c$ production, a neutrino scatters on the entire nucleus via the weak interaction, leaving the nucleus intact while producing a muon and a pion in the final state. The reaction
\[
\nu_{\mu} + \mathrm{Ar} \;\rightarrow\; \mu^{-} + \pi^{+} + \mathrm{Ar}
\]
is characterized by a forward-going lepton and meson and minimal vertex activity, since the target nucleus remains in its ground state. Theoretical descriptions of $\nu_\mu$CC$\pi^+_c$ production are rooted in the partially conserved axial current (PCAC) hypothesis \cite{ReinSehgal1983}. The Rein-Sehgal formalism \cite{ReinSehgal_Update} and later refinements by Berger and Sehgal \cite{BergerSehgal_2009} incorporate lepton-mass effects and improved treatments of pion-nucleus scattering. While alternative microscopic models of $\nu_\mu$CC$\pi^+_c$ production exist, the Berger-Sehgal model is commonly implemented in neutrino event generators and is used as the baseline prediction in this analysis.

Experimentally, CC coherent pion production has been measured on a variety of targets, including carbon, oxygen, lead, and iron, by MINERvA and T2K \cite{MINERvA_CohCC,T2K_CohCC, T2K_CohCC_2023}. Complementary measurements of neutral current coherent $\pi^0$ production have been reported by NOMAD,  MINOS, and NOvA, offering additional constraints on coherent interaction modeling \cite{KULLENBERG2009177, PhysRevD.94.072006, PhysRevD.102.012004}. The ArgoNeuT experiment provided the first measurement on argon at multi-GeV neutrino energies (with a mean of $\sim 4$~GeV) using a small LArTPC exposed to the NuMI beam \cite{ArgoNeuT2014}. Measurements at lower neutrino energies have not yet been reported. 

In this Letter, we present the first measurement of the flux-averaged $\nu_\mu$CC$\pi^+_c$ production cross section on argon at sub-GeV neutrino energies. The measurement is performed with the MicroBooNE  LArTPC \cite{MicroBooNE_Detector} exposed to the Fermilab Booster Neutrino Beam (BNB). The dataset corresponds to an exposure of $1.26\times10^{21}$ protons on target (POT) collected between 2015 and 2020. 

The MicroBooNE detector is a single-phase LArTPC with an active mass of 85\,t (metric tons) enclosed in a 170\,t cryostat, located 470 m downstream of the BNB target. Its active volume ($2.32\times2.56\times10.36 \, \text{m}^3$) is bounded by a cathode plane and three anode wire planes oriented at $0^{\circ}$ and $\pm60^{\circ}$ relative to the vertical \cite{MicroBooNE_Detector}. The fine-grained tracking and calorimetry of the detector, together with the large available statistics, enable isolation of this rare process, predicted to account for only about 0.15\% of neutrino interactions in simulation, and a detailed study of its kinematics providing quantitative input to neutrino-nucleus interaction models.

The BNB delivers a predominantly $\nu_{\mu}$ flux with a mean energy of 0.8~GeV, composed of 93.6\%  $\nu_{\mu}$, 5.86\% $\bar{\nu}_{\mu}$, and 0.57\% $\nu_e + \bar{\nu}_e$. In this energy regime, quasielastic (QE) interactions dominate and are topologically distinct from $\nu_\mu$CC$\pi^+_c$ production, making them relatively easy to reject. At higher neutrino energies, pion-rich resonant (RES) and deep-inelastic scattering (DIS) processes become more common and introduce backgrounds that are harder to reject, with little change in the $\nu_\mu$CC$\pi^+_c$ cross section. MicroBooNE’s millimeter-scale spatial resolution provides angular precision of $\sim$\,$2.4^{\circ}$ for forward muons, a capability essential for isolating the coherent topology from background processes.

\textit{Simulation and Reconstruction ---} The BNB flux is simulated with the MiniBooNE flux model adapted to the MicroBooNE detector location \cite{MiniBooNE_Flux}. The flux prediction is based on the Geant4 framework with external hadron production data providing constraints on $\pi^{\pm}$ and $K^{+}$ production \cite{HARP2007,SciBooNE2011}. Neutrino interactions in liquid argon are simulated using \texttt{GENIE v3.0.6} \cite{GENIE2010}
with the \texttt{G18\_10a\_02\_11a} model set. The MicroBooNE tune \cite{MicroBooNE_Tune} is applied to this baseline configuration, modifying the CCQE and CC2p2h (MEC) interaction models based on constraints from external data from T2K. The tune does not affect the modeling of RES, coherent, or DIS scattering processes. In \texttt{GENIE}, $\nu_\mu$CC$\pi^+_c$ is modeled using the Berger-Sehgal formalism \cite{BergerSehgal_2009}. Resonant pion production, which constitutes the dominant background to this measurement, is described using the Kuzmin-Lyubushkin-Naumov Berger-Sehgal model \cite{10.1063/1.3274164, doi:10.1142/S0217732304016172, ReinSehgal_Update, PhysRevD.77.053001}. Final-state interactions (FSI) are modeled using the hA2018 intranuclear cascade.

Particle propagation through the detector is simulated with \texttt{Geant4 v4\_10\_3\_p03c} using the \texttt{QGSP\_BERT} physics list, which accounts for hadronic interactions and electromagnetic processes in liquid argon \cite{Agostinelli2003, Allison2006, Allison2016}. The detector response to ionization electrons and scintillation light is modeled including diffusion, attenuation, and recombination effects \cite{Adams2018_SP1, Adams2018_SP2}. Simulated neutrino interactions are overlaid with data collected when the neutrino beam was off to reproduce backgrounds from electronics noise and cosmic rays, the latter being significant for a surface-based detector like MicroBooNE. Additional samples of simulated neutrino interactions outside the cryostat are included to account for “out-of-cryostat” backgrounds from upstream interactions. 

Event reconstruction is carried out using the Pandora multi-algorithm pattern recognition framework \cite{MicroBooNE_Pandora2018}. The same reconstruction chain is applied to both data and simulation. Following signal processing and hit formation from the TPC wires, Pandora clusters hits into candidate tracks and showers, builds three-dimensional topologies, and assigns parent-daughter relationships between reconstructed particles. Cosmic-ray rejection is performed using timing information from the photomultiplier tube system to distinguish beam-related activity from out-of-time cosmic rays. It also relies on topological classification of reconstructed interaction candidates identified by Pandora. Particle identification is applied using calorimetric and topological information including a log-likelihood ratio (LLR PID) method that compares measured $dE/dx$ profiles against particle hypotheses \cite{MicroBooNE_LLRPID2021}. 

\textit{Signal Definition and Event Selection ---} The signal is defined as $\nu_\mu$CC$\pi^+_c$ production on argon corresponding to interactions in which the final state contains exactly one muon and one charged pion, with no additional mesons or nucleons produced and with the nucleus remaining in its ground state. True signal events are identified using the coherent interaction classification provided by the \texttt{GENIE} event generator. The \texttt{GENIE} simulation is employed only to model the expected shapes of the signal and background distributions and to evaluate selection efficiencies. To ensure efficient reconstruction and selection, the analysis phase space is further restricted by requiring true muon and pion momenta of $p_{\mu},\,p_{\pi} > 150~\mathrm{MeV}/c$ and an opening angle $\theta_{\mu\pi} < 55^{\circ}$. The restricted signal phase space contains 92.2\% of the total simulated $\nu_\mu$CC$\pi^+_c$ production cross section. The reconstruction of muon and pion momenta is performed using multiple Coulomb scattering (MCS) \cite{Abratenko_2017}, as range-based momentum estimates are not reliable for tracks that exit the active detector volume.

Candidate events are required to have a reconstructed neutrino interaction vertex within the MicroBooNE fiducial volume ($10.0 \leq x \leq 246.4$~cm, $-106.5 \leq y \leq 106.5$~cm, $10.0 \leq z \leq 1000.0$~cm), where $x$, $y$, and $z$ denote the horizontal drift, vertical, and beam directions, respectively. They are further required to have exactly two well-reconstructed tracks originating within 10\,cm of the vertex, consistent with the muon-pion final state expected from $\nu_\mu$CC$\pi^+_c$ interactions. Tracks are required to be contained in the transverse ($x,y$) directions, while no containment is imposed on the tracks in the downstream beam ($z$) direction. The 77\% of simulated signal events have both tracks contained; the remaining 23\% have at least one track exiting the detector in the z-direction. This reflects the strongly forward-going nature of the coherent signal and improves overall selection efficiency and sample statistics. To remove events containing protons and highly ionizing tracks, both selected tracks are required to be consistent with \textit{minimally ionizing particles}, defined by a truncated mean $dE/dx < 2.5~\mathrm{MeV/cm}$ calculated using hits from the first third of the track, well before the Bragg peak \cite{abratenko2025measurementsinglechargedpion}, and to satisfy the LLR PID criterion that quantifies consistency with muon or pion hypotheses relative to a proton. To suppress the backgrounds from low-energy protons and poorly reconstructed tracks, both tracks are, additionally, required to have a reconstructed length greater than 20~cm, which ensures reliable calorimetric measurements and particle identification. These requirements ensure consistency with the expected signatures of a muon and a pion while rejecting tracks consistent with protons.

The forward, collimated topology of the coherent signal is isolated using angular requirements on the muon and pion. The opening angle between the two tracks is required to be less than $55^{\circ}$, which removes 82\% of the large-angle pion production from RES and DIS scattering. In addition, the cone angle ($\theta_{\text{cone}}$), defined as the angle between the direction of the neutrino beam and the momentum vector of the muon+pion system, is used as a discriminating observable between coherent and resonant pion production.

After all selections, the resulting sample achieves a signal purity of 32\% at an efficiency of 14.5\% in the signal-enhanced region, defined by $\theta_{\mathrm{cone}} < 16^\circ$, which is chosen to maximize signal purity while retaining reasonable efficiency. The remaining background is dominated by resonant pion production, which constitutes an irreducible background due to its similar final-state topology, particularly when the recoil nucleon is either below detection threshold or not reconstructed, including channels with final-state neutrons. The complementary region, $16^\circ < \theta_{\mathrm{cone}} < 56^\circ$, defines a background-dominated control region used to validate the background model. 
\begin{figure}[h!]
\includegraphics[width=0.95\linewidth]{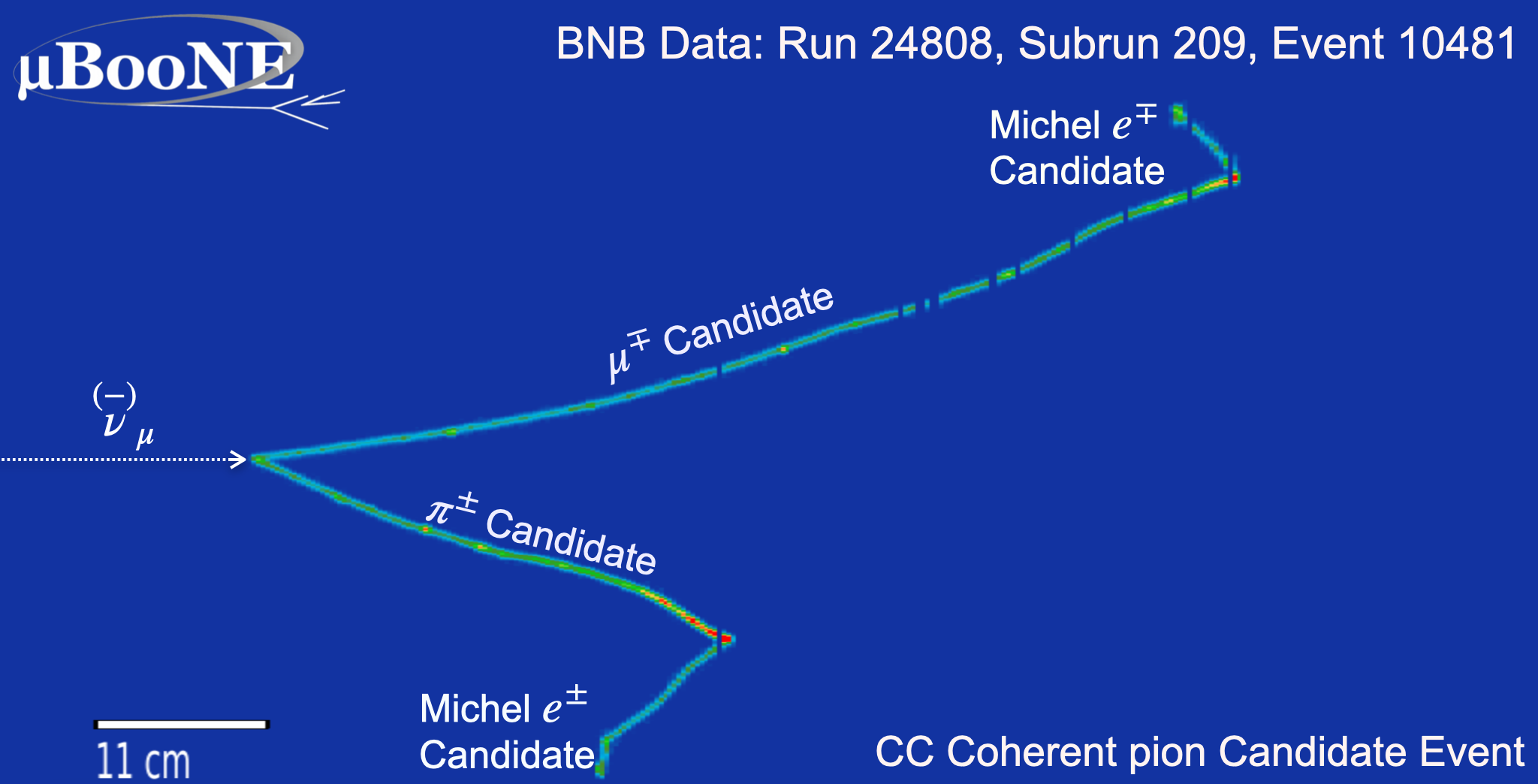}
\caption{\label{fig:eventdisplay}
Event display of a $\nu_\mu$CC$\pi^+_c$ candidate from BNB data.}
\end{figure}

Figure~\ref{fig:eventdisplay} shows the event display of a $\nu_\mu$CC$\pi^+_c$ candidate in MicroBooNE from BNB data. The reconstructed topology shows forward-going muon and pion candidates emerging from a common vertex with no visible additional tracks or vertex activity, consistent with the coherent scattering signature. The primary muon produces a visible Michel electron candidate, while the muon from the pion decay is not visible and is inferred from a Michel electron consistent with $\pi^{\pm} \to \mu^{\pm} \to e^{\pm}$.

\begin{figure}[h]
\includegraphics[width=1.0\linewidth]{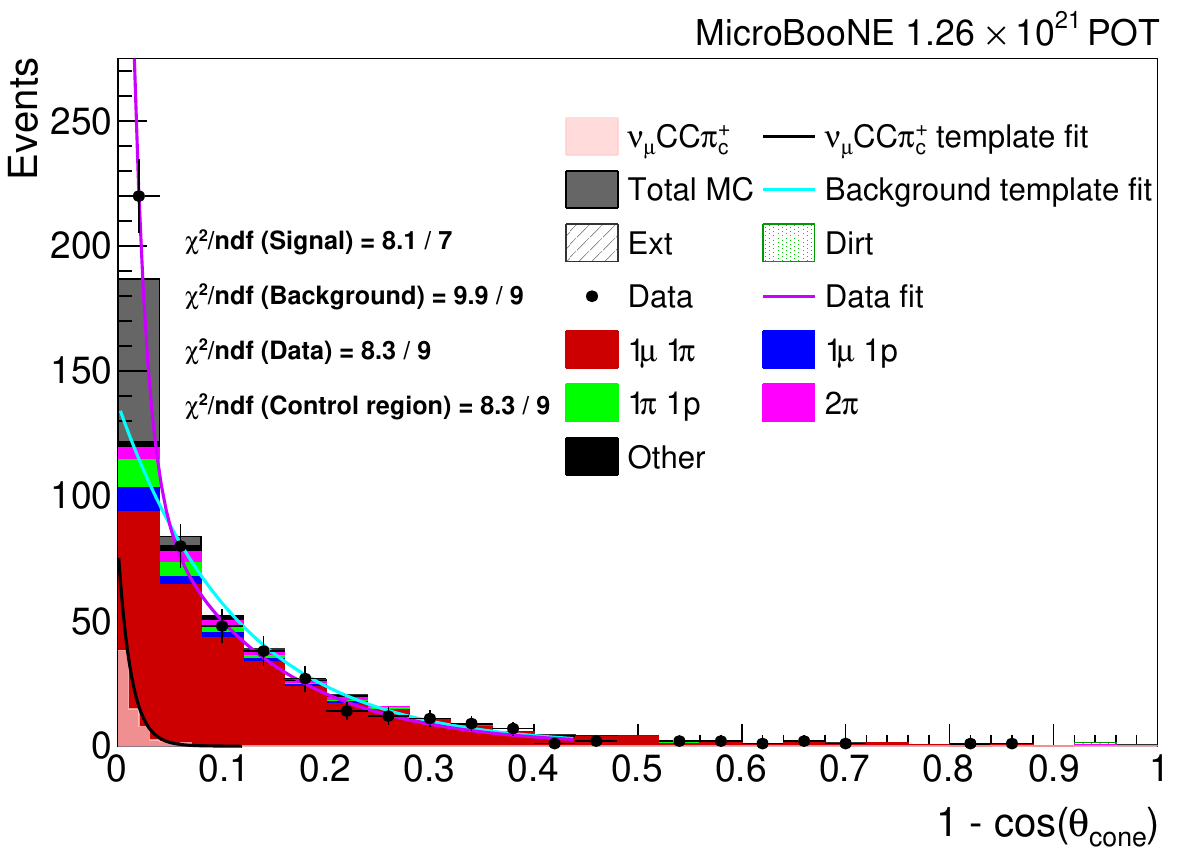}
\caption{\label{fig:coneangle_data}
Distribution of $1-\cos\theta_{\rm cone}$ for selected events in data compared with the total MC prediction and the fitted signal-plus-background model. The $\nu_\mu$CC$\pi^+_c$ signal exhibits a pronounced forward peak characteristic of coherent pion production, while non-coherent backgrounds form a smoothly falling spectrum at larger angles. The truth-selected signal prediction (pink) is overlaid with finer binning to illustrate the narrow forward peak and is not included in the stacked background. The $\chi^{2}/\mathrm{ndf}$ values for the signal-template fit, background-template fit, simultaneous fit to data, and control-region validation are reported on the plot.}
\end{figure}

\textit{Signal Extraction ---} The simulated $\nu_\mu$CC$\pi^+_c$ signal and background distributions are both forward-peaked, as shown in Fig.~\ref{fig:coneangle_data}; however, their angular distributions exhibit distinct behaviors. The signal produces a sharp, localized peak at small cone angles; non-coherent backgrounds show a smoothly falling spectrum. This difference in shape enables a data-driven separation of signal and background contributions. A finer binning is used for the simulated signal template to better resolve the coherent peak. Analytic parameterizations provide a stable representation of the simulated signal and background shapes while reducing sensitivity to bin-to-bin statistical fluctuations. The corresponding signal and background templates are modeled as exponential functions, $Y_{\text{Signal}}(\theta)=p_{0}\,e^{p_{1}\theta}$ and $Y_{\text{Background}}(\theta)=q_{0}\,e^{q_{1}\theta}$, where $p_{1}$ and $q_{1}$ describe the shape parameters extracted from simulation. The signal and background templates are first fit to simulated events to determine the shape parameters $p_1$ and $q_1$. The normalization constants $p_{0}$ and $q_{0}$ cancel when constructing the normalized templates used in Eq.~(\ref{eq:fit_model}). In the fit to data, $Y_{\text{Data}}(\theta)$, the shapes ($p_{1}$ and $q_{1}$) are fixed while the signal ($S$) and background ($B$) normalizations are allowed to float according to
\begin{equation}
Y_{\text{Data}}(\theta) = S \cdot \frac{Y_{\text{Signal}}(\theta)}{\int Y_{\text{Signal}}\,d\theta} 
                  + B \cdot \frac{Y_{\text{Background}}(\theta)}{\int Y_{\text{Background}}\,d\theta}.
\label{eq:fit_model}
\end{equation}
This formulation isolates the coherent-signal contribution from the background through a data-driven determination of their normalizations, while remaining largely insensitive to simulation-dependent rate uncertainties. It is validated using simulated events treated as data, reproducing signal and background normalizations within statistical uncertainties. This strategy exploits the distinct kinematic differences between coherent and non-coherent processes while minimizing reliance on simulated cross sections or event rates enabling a robust, data-driven extraction of the coherent signal yield.

After applying all the $\nu_\mu$CC$\pi^+_c$ selection criteria, a total of 464 events are observed in the data. The simultaneous fit to the data described above, performed using MINUIT \cite{minuit}, estimates $125 \pm 17.2_\mathrm{stat} \pm 13.7_\mathrm{syst}$ coherent signal events and $333 \pm 22.4_\mathrm{stat} \pm 14.6_\mathrm{syst}$ non-coherent background events. The quoted systematic uncertainties on the fitted signal and background yields include only template-shape variations propagated through the fit. The fit decomposes the observed distribution into signal-like and background-like components using template shapes determined from simulation. An excess of events is observed in the forward region ($\theta_{\mathrm{cone}} < 16^{\circ}$), where the coherent signal is expected to dominate. The \texttt{GENIE} simulation predicts $68$ signal events. While the fit uses the GENIE-predicted signal shape, its normalization is unconstrained, and the observed forward excess results in the fitted yield of 125 events. Fits to the simulated signal and background templates yield $\chi^{2}/\mathrm{ndf}=8.1/7$ and $9.9/9$, respectively. Using these fixed template shapes, the simultaneous fit to the data yields $\chi^{2}/\mathrm{ndf}=8.3/9$. A comparable value of $\chi^{2}/\mathrm{ndf}=8.3/9$ is obtained in the background-dominated control region, indicating good agreement between the fitted model and the observed data. Contributions from cosmic rays and out-of-cryostat interactions are negligible.

\begin{figure}[h]
\includegraphics[width=1.0\linewidth]{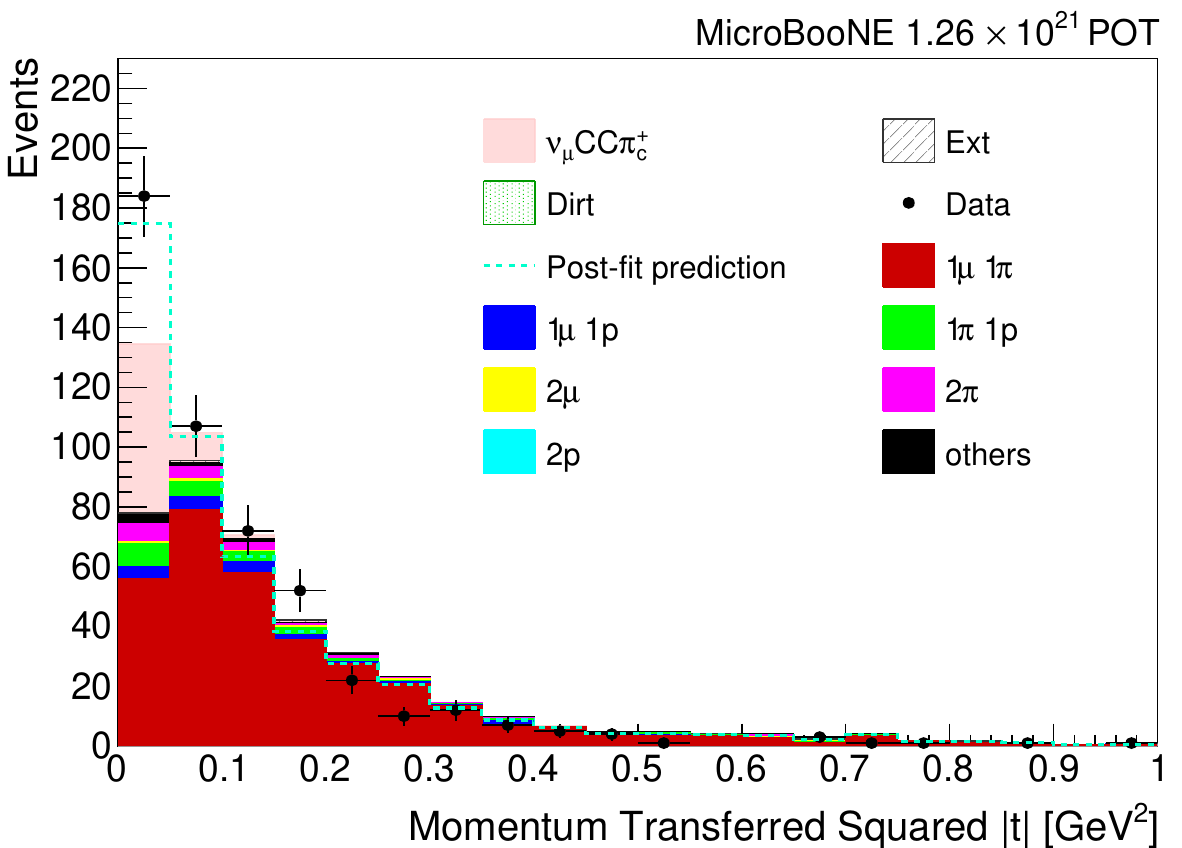}
\caption{\label{fig:q2}
Distribution of the squared four-momentum transfer, $|t|$. The blue dashed line shows the post-fit prediction obtained using the fitted signal and background normalizations, while the colored stacked histograms correspond to the nominal \texttt{GENIE} simulation. The selected sample exhibits the expected suppression at large $|t|$, consistent with the presence of a coherent signal.}
\end{figure}

As an additional validation of $\nu_\mu$CC$\pi^+_c$ production, we consider the squared four-momentum transfer to the argon nucleus, $|t| \equiv |(p_\nu - p_\mu - p_{\pi})^2|$, reconstructed from the measured muon and pion four-momenta. The post-fit $|t|$ distribution shown in Fig.~\ref{fig:q2} is obtained by applying the fitted signal ($S$) and background ($B$) normalizations from the $1-\cos\theta_{\text{cone}}$ fit to the simulated $|t|$ templates. This procedure propagates the fit results without introducing additional degrees of freedom. The resulting distribution exhibits an excess at low $|t|$, as expected for coherent scattering, providing an independent validation of the extracted signal.

\textit{Systematic Uncertainties ---} Systematic uncertainties are evaluated by propagating variations in the detector response, neutrino flux, hadronic reinteraction modeling, and neutrino interaction cross-section modeling through the full data-driven fitting procedure. Each category is independently assessed, and the resulting variations in the normalizations of the fitted signal ($S$) and the background ($B$) are combined to form the total systematic uncertainty.

Detector-related uncertainties are evaluated using dedicated simulation samples in which individual detector parameters are varied within their estimated uncertainties. The varied detector response parameters include light attenuation and yield in liquid argon, optical scattering effects \cite{microboonecollaboration2026scintillationlightcalibrationssystematic}, distortions in the electric and wire-field response, variations in the ion-electron recombination model, and space-charge effects that modify the reconstructed charge distribution \cite{Abratenko_2020, Abratenko2022UNISIM}. These variations collectively account for uncertainties in charge and light collection, which impact energy and $dE/dx$ reconstruction.

Uncertainties associated with the BNB flux \cite{MicroBooNE2018BNBFlux}, \texttt{GENIE} cross-section modeling \cite{MicroBooNE_Tune}, and hadronic reinteractions \cite{Calcutt_2021} in the detector are evaluated by varying the relevant parameters and propagating these variations through the analysis. Ensembles of systematically varied simulation samples are generated to capture correlated effects arising from hadron production, neutrino interaction models, and intranuclear transport processes.

The central value (CV) fit to simulation provides the shape parameters $p_{1}^{CV}$ and $q_{1}^{CV}$ for signal and background, respectively, obtained from MINUIT \cite{minuit}. This central value fit is performed once and serves as the reference for all systematic variations. Systematic uncertainties are evaluated by repeating the fits on systematically varied samples. For each variation, new shape parameters for signal and background are obtained, and their shifts relative to the CV fit are defined as $\delta{p}$ and $\delta{q}$. A fit to the data is repeated every time for $p_{1}^{CV} \pm \delta{p}$ and $q_{1}^{CV} \pm \delta{q}$ separately. The resulting change in $S$ and $B$ is recorded as systematic error due to that specific variation.

The total systematic uncertainty on the $\nu_\mu$CC$\pi^+_c$ cross section is 13.1\%, obtained by combining in quadrature the systematic uncertainty on the fitted signal yield with normalization uncertainties from the neutrino flux, POT determination, and number of argon targets. The dominant contributions arise from neutrino flux prediction (8.1\%), cross-section modeling (7.5\%), and detector effects (6.6\%). The cross-section modeling uncertainty includes variations of the Berger-Sehgal coherent-pion axial mass ($40\%$) and effective nuclear radius ($10\%$), which change the extracted cross section by 6.0\% and 0.4\%, respectively, together with a 4.5\% contribution from variations of the background shape. Subdominant contributions come from the POT determination (2.0\%), hadronic reinteractions (1.2\%), and the number of argon targets (1.0\%).

The cross section extraction procedure is validated using simulated events generated with an alternative interaction model and treated as data. In particular, a NuWro sample, which predicts significantly different $\nu_\mu$CC$\pi^+_c$ production rates compared to \texttt{GENIE}, is used to test the robustness of the method. The extracted signal and background yields are consistent with the corresponding truth values within statistical uncertainties, demonstrating that the largely normalization-independent approach is robust against such model differences.

\textit{Results ---} The flux-averaged cross section is obtained from the fitted signal yield as
\begin{equation}
\sigma_{\mathrm{CC\,Coh}} = \frac{N_{\mathrm{sig}}} {\epsilon\,N_{\mathrm{Ar}}\,\Phi}
\label{eq:xsec}
\end{equation}
where $N_{\mathrm{sig}}$ is the fitted signal yield, $\epsilon$ is the signal-selection efficiency, $N_{\mathrm{Ar}}$ is the number of argon target nuclei in the fiducial volume, and $\Phi$ is the integrated $\nu$ flux corresponding to the analyzed exposure.
The resulting total flux-averaged $\nu_\mu$CC$\pi^+_c$ production cross section on argon is measured to be
\begin{equation*}
    \sigma_{\text{CC\,Coh}} = (9.1 \pm 1.2 _{\text{stat}} \pm 1.2_\text{syst}) \times 10^{-40}\,\text{cm}^{2}/\text{Ar.}
\end{equation*}

This result represents the first measurement of the $\nu_\mu$CC$\pi^+_c$ production cross section on an argon target at sub-GeV neutrino energies. The measurement is compared with model predictions implemented in the \texttt{GENIE 3.0.6 G18\_10a\_02\_11a} \cite{MicroBooNE_Tune} (labeled \texttt{GENIE 3.0.6 G18 $\mu B$}), \texttt{NEUT 6.1.4} \cite{Hayato2021NEUT}, and \texttt{NuWro 25.11} \cite{Golan2012FSI} neutrino event generators, all based on the same Berger-Sehgal (BS) formalism. The comparison also includes an alternative \texttt{GENIE 3.6.2 G18\_10a\_02\_11a} configuration based on the Rein-Sehgal (RS) model. The comparison is performed using the \texttt{NUISANCE} framework \cite{Stowell_2017}.

A summary of the measured and predicted cross sections is presented in Table~\ref{tab:coh_results}. The measured cross section is in good agreement with the \texttt{NEUT} and \texttt{GENIE} RS predictions, while the \texttt{GENIE} Berger-Sehgal (BS) and \texttt{NuWro} predictions show increasing tension with the data. Using the combined statistical and systematic uncertainty, a one-bin ($\mathrm{ndf}=1$) $\chi^2$ comparison yields $\chi^2 = 0.3$ for \texttt{GENIE} RS, $\chi^2 = 0.9$ for \texttt{NEUT}, $\chi^2 = 5.8$ for \texttt{GENIE} BS, and $\chi^2 = 8.3$ for \texttt{NuWro}. The spread in predictions reflects differences in the underlying implementations of $\nu_\mu$CC$\pi^+_c$ production across generators. Discrepancies between the data and several generators, as well as among generators themselves, highlight the need for precise cross-section measurements of this process.
\begin{table}[h]
\centering
\caption{Comparison of the measured $\nu_\mu$CC$\pi^+_c$ cross section on argon with model predictions. All values are flux-averaged and expressed in units of $10^{-40}\,\text{cm}^2/\text{Ar}$.}
\label{tab:coh_results}
\begin{ruledtabular}
\begin{tabular}{l c c}
Model/Measurement & Cross Section & \textbf{$\chi^2/n_{bins}$}\\
\hline
Measured (MicroBooNE) & \multicolumn{2}{l}{$9.1 \pm 1.2_{\text{stat}} \pm 1.2_\text{syst}$} \\ \hline
\scriptsize \texttt{GENIE 3.0.6 G18 $\mu B$} & $5.0$ & $5.8/1$ \\
\scriptsize \texttt{GENIE 3.6.2 G18\_10a\_02\_11a} (RS) & $10.1$ & $0.3/1$ \\
\scriptsize \texttt{NEUT 6.1.4}  & $10.7$ & 0.9/1\\
\scriptsize \texttt{NuWro 25.11}  & $14.0$ & 8.3/1\\
\end{tabular}
\end{ruledtabular}
\end{table}

\textit{Conclusion ---} We have presented the first measurement of the flux-averaged $\nu_\mu$CC$\pi^+_c$ production cross section on argon at sub-GeV neutrino energies using the MicroBooNE detector exposed to the Fermilab Booster Neutrino Beam.  This analysis isolates the coherent topology by exploiting the fine angular resolution of a liquid-argon time projection chamber and employs a data-driven fit to extract the signal yield with minimal model dependence. The measured cross section is consistent with \texttt{NEUT} and \texttt{GENIE} Rein-Sehgal predictions, while \texttt{GENIE} Berger-Sehgal and \texttt{NuWro} show tension with the data. This result provides the most precise constraint on this process to date at BNB energies and establishes a benchmark for improving neutrino-nucleus interaction models. Such measurements are useful for reducing flux uncertainties in future long-baseline experiments like DUNE, where coherent pion production has been proposed as a standard candle for $\nu_{\mu}$ flux normalization. Achieving the few-percent precision required for coherent pion production to serve as a neutrino flux constraint will require substantially larger event samples and improved control of interaction-model and detector systematic uncertainties. The larger samples anticipated from SBND \cite{SBND} and ICARUS \cite{ICARUS_SBN} offer the potential to advance toward this level of precision.

\textit{Acknowledgements ---} This document was prepared by the MicroBooNE collaboration using the resources of the Fermi National Accelerator Laboratory (Fermilab), a U.S. Department of Energy, Office of Science, Office of High Energy Physics HEP User Facility. Fermilab is managed by Fermi Forward Discovery Group, LLC, acting under Contract No. 89243024CSC000002. MicroBooNE is supported by the
following: 
the U.S. Department of Energy, Office of Science, Offices of High Energy Physics and Nuclear Physics; 
the U.S. National Science Foundation; 
the Swiss National Science Foundation; 
the Science and Technology Facilities Council (STFC), part of United Kingdom Research and Innovation (UKRI);
the Royal Society (United Kingdom);
the UKRI Future Leaders Fellowship;
the NSF AI Institute for Artificial Intelligence and Fundamental Interactions;
and the European Union’s Horizon 2020 research and innovation programme under the Marie Sk\l{}odowska-Curie grant agreement No. 101003460 (PROBES). Additional support for the laser calibration system and cosmic ray tagger was provided by the Albert Einstein Center for Fundamental Physics, Bern, Switzerland. We also acknowledge the contributions of technical and scientific staff to the design, construction, and operation of the MicroBooNE detector as well as the contributions of past collaborators to the development of MicroBooNE analyses, without whom this work would not have been possible.

\bibliography{main}% Produces the bibliography via BibTeX.

@article{Abi2020DUNE,
  author        = {Abi, B. and others},
  collaboration = {DUNE Collaboration},
  title         = {{Long-baseline neutrino oscillation physics potential of the DUNE experiment}},
  journal       = {Eur. Phys. J. C},
  volume        = {80},
  pages         = {978},
  year          = {2020},
  doi           = {https://doi.org/10.1140/epjc/s10052-020-08456-z}
}

@article{Jung2025,
  author       = {Mun Jung Jung and Vishvas Pandey and Gray Putnam and David W. Schmitz},
  title        = {{Neutrino Induced Charged Current Coherent Pion Production for Constraining the Muon Neutrino Flux at DUNE}},
  journal      = {arXiv preprint},
  eprint       = {2502.02576},
  year         = {2025},
  archivePrefix= {arXiv},
  primaryClass = {hep-ph}
}

@article{Adler,
  title = {{Tests of the Conserved Vector Current and Partially Conserved Axial-Vector Current Hypotheses in High-Energy Neutrino Reactions}},
  author = {Adler, Stephen L.},
  journal = {Phys. Rev.},
  volume = {135},
  issue = {4B},
  pages = {B963--B966},
  numpages = {0},
  year = {1964},
  month = {Aug},
  publisher = {American Physical Society},
  doi = {10.1103/PhysRev.135.B963},
  url = {https://link.aps.org/doi/10.1103/PhysRev.135.B963}
}

@article{ReinSehgal1983,
  author       = {D. Rein and L. M. Sehgal},
  title        = {{Coherent $\pi^0$ production in neutrino reactions}},
  journal      = {Nucl. Phys. B},
  volume       = {223},
  pages        = {29--44},
  year         = {1983},
  doi          = {10.1016/0550-3213(83)90090-1}
}

@article{ReinSehgal_Update,
  author        = {C. Berger and L. M. Sehgal},
  title         = {Lepton mass effects in single pion production by neutrinos},
  journal       = {Phys. Rev. D},
  volume        = {76},
  pages         = {113004},
  year          = {2007},
  doi           = {10.1103/PhysRevD.76.113004}
}

@article{BergerSehgal_2009,
  author        = {C. Berger and L. M. Sehgal},
  title         = {Partially conserved axial vector current and coherent pion production by low energy neutrinos},
  journal       = {Phys. Rev. D},
  volume        = {79},
  pages         = {053003},
  year          = {2009},
  doi           = {10.1103/PhysRevD.79.053003}
}

@article{PhysRevD.77.053001,
  title = {Form factors in the quark resonance model},
  author = {{K. M. Graczyk and J. T. Sobczyk}},
  journal = {{Phys. Rev. D}},
  volume = {77},
  issue = {5},
  pages = {053001},
  numpages = {12},
  year = {2008},
  month = {Mar},
  publisher = {{American Physical Society}},
  doi = {10.1103/PhysRevD.77.053001},
  url = {https://link.aps.org/doi/10.1103/PhysRevD.77.053001}
}

@article{MINERvA_CohCC,
 title = {{Neutrino-Induced Coherent ${\ensuremath{\pi}}^{+}$ Production in C, CH, Fe, and Pb at $⟨{E}_{\ensuremath{\nu}}⟩\ensuremath{\sim}6\text{ }\text{ }\mathrm{GeV}$}},
  author = {Ram\'{\i}rez, M. A. and others},
  collaboration = {{MINERvA Collaboration}},
  journal = {{Phys. Rev. Lett.}},
  volume = {131},
  issue = {5},
  pages = {051801},
  numpages = {8},
  year = {2023},
  month = {Aug},
  publisher = {{American Physical Society}},
  doi = {10.1103/PhysRevLett.131.051801},
  url = {https://link.aps.org/doi/10.1103/PhysRevLett.131.051801}
}

@article{T2K_CohCC,
  author        = {K. Abe and others },
  collaboration = {T2K Collaboration},
  title         = {First measurement of the $\nu_\mu$ charged-current single $\pi^+$ cross section on water with the {T2K} near detector},
  journal       = {Phys. Rev. D},
  volume        = {95},
  pages         = {012010},
  year          = {2017},
  doi           = {10.1103/PhysRevD.95.012010}
}

@article{T2K_CohCC_2023,
  title = {Measurements of the ${\ensuremath{\nu}}_{\ensuremath{\mu}}$ and ${\overline{\ensuremath{\nu}}}_{\ensuremath{\mu}}$-induced coherent charged pion production cross sections on $^{12}\mathrm{C}$ by the {T2K} experiment},
  author = {K. Abe and others},
  collaboration = {T2K Collaboration},
  journal = {Phys. Rev. D},
  volume = {108},
  issue = {9},
  pages = {092009},
  numpages = {15},
  year = {2023},
  month = {Nov},
  publisher = {American Physical Society},
  doi = {10.1103/PhysRevD.108.092009},
  url = {https://link.aps.org/doi/10.1103/PhysRevD.108.092009}
}

@article{ArgoNeuT2014,
  title = {{First Measurement of Neutrino and Antineutrino Coherent Charged Pion Production on Argon}},
  author = {Acciarri, R. and others},
  collaboration = {ArgoNeuT Collaboration},
  journal = {Phys. Rev. Lett.},
  volume = {113},
  issue = {26},
  pages = {261801},
  numpages = {4},
  year = {2014},
  month = {Dec},
  publisher = {American Physical Society},
  doi = {10.1103/PhysRevLett.113.261801},
  url = {https://link.aps.org/doi/10.1103/PhysRevLett.113.261801}
}

@article{MicroBooNE_Detector,
  author        = {Acciarri, R. and others},
  collaboration = {MicroBooNE Collaboration},
  title         = {{Design and Construction of the {MicroBooNE} Detector}},
  journal       = {J. Instrum.},
  volume        = {12},
  pages         = {P02017 (2017)},
  year          = {2017},
  doi           = {10.1088/1748-0221/12/02/P02017}
}

@article{MiniBooNE_Flux,
author = {A. A. Aguilar-Arevalo and others},
title = {{The Neutrino Flux Prediction at MiniBooNE}},
collaboration = {MiniBooNE Collaboration},
  journal = {Phys. Rev. D},
  volume = {79},
  issue = {7},
  pages = {072002},
  numpages = {38},
  year = {2009},
  month = {Apr},
  publisher = {American Physical Society},
  doi = {10.1103/PhysRevD.79.072002},
  url = {https://link.aps.org/doi/10.1103/PhysRevD.79.072002}
}

@article{HARP2007,
  author       = {Catanesi, M. G. and others},
  collaboration = {HARP Collaboration},
  title        = {Measurement of the production cross-section of positive pions in the collision of {8.9-GeV/\textit{c}} protons on beryllium},
  journal      = {Eur. Phys. J. C},
  volume       = {52},
  pages        = {29-53},
  year         = {2007},
  doi          = {https://doi.org/10.1140/epjc/s10052-007-0382-8}
}

@article{SciBooNE2011,
  author       = {G. Cheng and others},
  collaboration = {SciBooNE Collaboration},
  title        = {{Measurement of $K^+$ production cross section by 8 {GeV} protons using high energy neutrino interactions in the {SciBooNE} detector}},
  journal      = {Phys. Rev. D},
  volume       = {84},
  pages        = {012009},
  year         = {2011},
  doi          = {10.1103/PhysRevD.84.012009}
}

@article{GENIE2010,
  author       = {C. Andreopoulos and others},
  title        = {{The GENIE neutrino Monte Carlo generator}},
  journal      = {Nucl. Instrum. Meth. A},
  volume       = {614},
  pages        = {87},
  year         = {2010},
  doi          = {10.1016/j.nima.2009.12.009}
}

@article{MicroBooNE_Tune,
  author       = {P. Abratenko and others},
collaboration = {MicroBooNE Collaboration},
  title        = {{New CC0$\pi$ GENIE model tune for {MicroBooNE}}},
  journal      = {Phys. Rev. D},
  volume       = {105},
  pages        = {072001},
  year         = {2022},
  doi          = {10.1103/PhysRevD.105.072001}
}

@article{MicroBooNE_Pandora2018,
 author       = {R. Acciarri and others},
collaboration = {MicroBooNE Collaboration},
  title        = {{The Pandora multi-algorithm approach to automated pattern recognition of cosmic-ray muon and neutrino events in the {MicroBooNE} detector}},
  journal      = {Eur. Phys. J. C},
  volume       = {78},
  pages        = {82},
  year         = {2018},
  doi          = {10.1140/epjc/s10052-017-5481-6}
}

@article{MicroBooNE_LLRPID2021,
author       = {P. Abratenko and others},
collaboration = {MicroBooNE Collaboration},
  title        = {{Calorimetric classification of track-like signatures in liquid argon TPCs using {MicroBooNE} data}},
  journal      = {J. High Energy Phys.},
  volume       = {12},
  pages        = {153 (2021)},
  year         = {2021},
  doi          = {10.1007/JHEP12(2021)153}
}

@article{Agostinelli2003,
  author       = {S. Agostinelli and others},
  collaboration = {GEANT4 Collaboration},
  title        = {GEANT4---A simulation toolkit},
  journal      = {Nucl. Instrum. Meth. A},
  volume       = {506},
  pages        = {250},
  year         = {2003},
  doi          = {10.1016/S0168-9002(03)01368-8}
}

@article{Allison2006,
  author       = {J. Allison and others},
  title        = {Geant4 developments and applications},
  journal      = {IEEE Trans. Nucl. Sci.},
  volume       = {53},
  pages        = {270},
  year         = {2006},
  doi          = {10.1109/TNS.2006.869826}
}

@article{Allison2016,
  author       = {J. Allison and others},
  title        = {{Recent developments in Geant4}},
  journal      = {Nucl. Instrum. Meth. A},
  volume       = {835},
  pages        = {186},
  year         = {2016},
  doi          = {10.1016/j.nima.2016.06.125}
}

@article{Adams2018_SP1,
  author       = {C. Adams and others},
collaboration = {MicroBooNE Collaboration},
  title        = {{Ionization electron signal processing in single phase LArTPCs. Part I. Algorithm description and quantitative evaluation with {MicroBooNE} simulation}},
  journal      = {J. Instrum.},
  volume       = {13},
  pages        = {P07006 (2018)},
  year         = {2018},
  doi          = {10.1088/1748-0221/13/07/P07006}
}

@article{Adams2018_SP2,
 author       = {C. Adams and others},
collaboration = {MicroBooNE Collaboration},
  title        = {{Ionization electron signal processing in single phase LArTPCs. Part II. Data/simulation comparison and performance in {MicroBooNE}}},
  journal      = {J. Instrum.},
  volume       = {13},
  pages        = {P07007 (2018)},
  year         = {2018},
  doi          = {10.1088/1748-0221/13/07/P07007}
}

@ARTICLE{minuit,
  author={Hatlo, M. and James, F. and Mato, P. and Moneta, L. and Winkler, M. and Zsenei, A.},
  journal={{IEEE Trans. Nucl. Sci.}}, 
  title={{Developments of mathematical software libraries for the LHC experiments}}, 
  year={2005},
  volume={52},
  number={6},
  pages={2818-2822},
  doi={10.1109/TNS.2005.860152}}

@PREAMBLE{
 "\providecommand{\noopsort}[1]{}" 
 # "\providecommand{\singleletter}[1]{#1}%" 
}

@article{abratenko2025measurementsinglechargedpion,
  title = {{Measurement of single charged pion production in charged-current ${\ensuremath{\nu}}_{\ensuremath{\mu}}$-Ar interactions with the MicroBooNE detector}},
  author = {Abratenko, P. and others},
  collaboration = {MicroBooNE Collaboration},
  journal = {Phys. Rev. D},
  volume = {113},
  issue = {3},
  pages = {032007},
  numpages = {18},
  year = {2026},
  month = {Feb},
  publisher = {American Physical Society},
  doi = {10.1103/t2cw-cdx2},
  url = {https://link.aps.org/doi/10.1103/t2cw-cdx2}
}

@article{10.1063/1.3274164,
    author = {Nowak, Jaroslaw A. and others},
    collaboration = {{MiniBooNE Collaboration}},
    title = {{Four Momentum Transfer Discrepancy in the Charged Current $\pi^+$; Production in the MiniBooNE: Data vs. Theory}},
    journal = {{AIP Conf. Proc.}},
    volume = {1189},
    number = {1},
    pages = {243-248},
    year = {2009},
    month = {11},
    issn = {0094-243X},
    doi = {10.1063/1.3274164},
    url = {https://doi.org/10.1063/1.3274164},
}

@article{doi:10.1142/S0217732304016172,
author = {{K. S. Kuzmin, V. V. Lyubushkin, and V. A. Naumov}},
title = {{Lepton
polarization in neutrino-nucleon interactions}},
journal = {{Modern Physics Letters A}},
volume = {19},
number = {38},
pages = {2815-2829},
year = {2004},
doi = {10.1142/S0217732304016172},
URL = {https://doi.org/10.1142/S0217732304016172
}
}

@article{Hayato2021NEUT,
  author    = {{Y. Hayato and L. Pickering}},
  title     = {{The {NEUT} neutrino interaction simulation program library}},
  journal   = {{Eur. Phys. J. Spec. Top.}},
  volume    = {230},
  pages     = {4469-4481},
  year      = {2021},
  doi       = {10.1140/epjs/s11734-021-00287-7},
  url       ={https://doi.org/10.1140/epjs/s11734-021-00287-7}
}

@article{Golan2012FSI,
  author  = {{T. Golan, C. Juszczak and J. T. Sobczyk}},
  title   = {Final state interactions effects in neutrino-nucleus interactions},
  journal = {{Phys. Rev. C}},
  volume  = {86},
  pages   = {015505},
  year    = {2012},
  doi     = {10.1103/PhysRevC.86.015505},
  url     = {https://doi.org/10.1103/PhysRevC.86.015505}
}

@techreport{MicroBooNE2018BNBFlux,
  author      = {{MicroBooNE Collaboration}},
  title       = {{Booster Neutrino Flux Prediction at {MicroBooNE}}},
  institution = {Fermilab},
  note        = {\href{https://www.osti.gov/servlets/purl/1573216/}{MICROBOONE-NOTE-1031-PUB}},
  year        = {2018}
}

@article{Calcutt_2021,
doi = {10.1088/1748-0221/16/08/P08042},
url = {https://doi.org/10.1088/1748-0221/16/08/P08042},
year = {2021},
month = {aug},
publisher = {{IOP Publishing}},
volume = {16},
number = {08},
pages = {P08042},
author = {{J. Calcutt and C. Thorpe and K. Mahn and L. Fields}},
title = {{Geant4Reweight: A framework for evaluating and propagating hadronic interaction uncertainties in Geant4}},
journal = {{J. Instrum.}}
}

@article{Abratenko_2020,
doi = {10.1088/1748-0221/15/12/P12037},
url = {https://doi.org/10.1088/1748-0221/15/12/P12037},
year = {2020},
month = {dec},
publisher = {},
volume = {15},
number = {12},
pages = {P12037},
author = {Abratenko, P. and others},
collaboration = {MicroBooNE Collaboration},
title = {{Measurement of space charge effects in the MicroBooNE LArTPC using cosmic muons}},
journal = {{J. Instrum.}}
}

@article{Abratenko2022UNISIM,
  author        = {Abratenko, P. and others},
  collaboration = {MicroBooNE Collaboration},
  title         = {{Novel Approach for Evaluating Detector-Related Uncertainties in a {LArTPC} Using {MicroBooNE} Data}},
  journal       = {{Eur. Phys. J. C}},
  volume        = {82},
  pages         = {454},
  year          = {2022},
  doi           = {10.1140/epjc/s10052-022-10270-8}
}

@article{Stowell_2017,
doi = {10.1088/1748-0221/12/01/P01016},
url = {https://doi.org/10.1088/1748-0221/12/01/P01016},
year = {2017},
month = {jan},
publisher = {},
volume = {12},
pages = {P01016 (2017)},
author = {Stowell, P. and others},
title = {{NUISANCE: a neutrino cross-section generator tuning and comparison framework}},
journal = {J. Instrum.}
}

@article{Abratenko_2017,
doi = {10.1088/1748-0221/12/10/P10010},
url = {https://doi.org/10.1088/1748-0221/12/10/P10010},
year = {2017},
month = {oct},
publisher = {},
volume = {12},
number = {10},
pages = {P10010},
author = {Abratenko, P. and others},
title = {{Determination of muon momentum in the MicroBooNE LArTPC using an improved model of multiple Coulomb scattering}},
journal = {{J. Instrum.}}
}

@article{SBND,
  author        = {Acciarri, R. and others},
  collaboration = {SBND Collaboration},
  title         = {{The Short-Baseline Near Detector at Fermilab}},
  eprint        = {2504.00245},
  archivePrefix = {arXiv},
  primaryClass  = {hep-ex},
  journal = {arXiv preprint},
  year          = {2025}
}

@article{ICARUS_SBN,
doi = {10.1088/1748-0221/19/04/C04061},
url = {https://doi.org/10.1088/1748-0221/19/04/C04061},
year = {2024},
month = {apr},
publisher = {{IOP Publishing}},
volume = {19},
number = {04},
pages = {C04061},
author = {Torretta, D. and others},
collaboration = {ICARUS Collaboration},
title = {{Status and perspectives of the ICARUS experiment at the Fermilab Short Baseline Neutrino program}},
journal = {{J. Instrum.}}
}

@article{KULLENBERG2009177,
title = {{A measurement of coherent neutral pion production in neutrino neutral current interactions in the NOMAD experiment}},
journal = {{Phys. Lett. B}},
volume = {682},
number = {2},
pages = {177-184},
year = {2009},
issn = {0370-2693},
doi = {https://doi.org/10.1016/j.physletb.2009.10.083},
url = {https://www.sciencedirect.com/science/article/pii/S0370269309012908},
author = {C.T. Kullenberg and others}
}

@article{PhysRevD.94.072006,
  title = {{Measurement of single ${\ensuremath{\pi}}^{0}$ production by coherent neutral-current $\ensuremath{\nu}$ Fe interactions in the MINOS Near Detector}},
  author = {Adamson, P. and others},
  collaboration = {{MINOS Collaboration}},
  journal = {{Phys. Rev. D}},
  volume = {94},
  issue = {7},
  pages = {072006},
  numpages = {20},
  year = {2016},
  month = {Oct},
  publisher = {{American Physical Society}},
  doi = {10.1103/PhysRevD.94.072006},
  url = {https://link.aps.org/doi/10.1103/PhysRevD.94.072006}
}

@article{PhysRevD.102.012004,
  title = {{Measurement of neutrino-induced neutral-current coherent ${\ensuremath{\pi}}^{0}$ production in the NOvA near detector}},
  author = {Acero, M. A. and others},
  collaboration = {{NOvA Collaboration}},
  journal = {Phys. Rev. D},
  volume = {102},
  issue = {1},
  pages = {012004},
  numpages = {11},
  year = {2020},
  month = {Jul},
  publisher = {{American Physical Society}},
  doi = {10.1103/PhysRevD.102.012004},
  url = {https://link.aps.org/doi/10.1103/PhysRevD.102.012004}
}

@article{microboonecollaboration2026scintillationlightcalibrationssystematic,
      title={{Scintillation light calibrations, systematic uncertainties, and triggering efficiency in the MicroBooNE detector}}, 
      author = {P. Abratenko and others},
      collaboration = {{MicroBooNE Collaboration}},
      year={2026},
      journal = {arXiv preprint},
      eprint={2603.23691},
      archivePrefix={arXiv},
      primaryClass={physics.ins-det},
      url={https://arxiv.org/abs/2603.23691}, 
}

@article{PhysRevD.79.013002,
  title = {{Theoretical study of neutrino-induced coherent pion production off nuclei at T2K and MiniBooNE energies}},
  author = {Amaro, J. E. and Hern\'andez, E. and Nieves, J. and Valverde, M.},
  journal = {Phys. Rev. D},
  volume = {79},
  issue = {1},
  pages = {013002},
  numpages = {18},
  year = {2009},
  month = {Jan},
  publisher = {American Physical Society},
  doi = {10.1103/PhysRevD.79.013002},
  url = {https://link.aps.org/doi/10.1103/PhysRevD.79.013002}
}

\end{document}